\documentclass[journal=jacsat,manuscript=article]{achemso}

\usepackage{chemformula} 
\usepackage[T1]{fontenc} 
\usepackage{booktabs}
\usepackage{graphicx}
\usepackage{subcaption}
\usepackage{siunitx}
\usepackage{float}     
\usepackage{placeins}  
\usepackage{subcaption} 

\author{Rishabh Dey}
\email{rishabh3dey@gmail.com}
\author{Michael Brocidiacono}
\author{Kushal Koirala}
\author{Alexander Tropsha}
\author{Konstantin I. Popov}
\email{kpopov@unc.edu}

\affiliation[University of North Carolina At Chapel Hill]
{Eshelman School of Pharmacy, University of North Carolina at Chapel Hill}

\title[An \textsf{achemso} demo]
  {Extending machine learning model for implicit solvation to free energy calculations}

\abbreviations{IR,NMR,UV}

\begin{document}

\begin{tocentry}

Some journals require a graphical entry for the Table of Contents.
This should be laid out ``print ready'' so that the sizing of the
text is correct.

Inside the \texttt{tocentry} environment, the font used is Helvetica
8\,pt, as required by \emph{Journal of the American Chemical
Society}.

The surrounding frame is 9\,cm by 3.5\,cm, which is the maximum
permitted for  \emph{Journal of the American Chemical Society}
graphical table of content entries. The box will not resize if the
content is too big: instead it will overflow the edge of the box.

This box and the associated title will always be printed on a
separate page at the end of the document.

\end{tocentry}

\begin{abstract}
  The implicit solvent approach offers a computationally efficient framework to model solvation effects in molecular simulations. However, its accuracy often falls short compared to explicit solvent models,  limiting its use in precise thermodynamic calculations. Recent advancements in machine learning (ML) present an opportunity to overcome these limitations by leveraging neural networks to develop more precise implicit solvent potentials for diverse applications. A major drawback of current ML-based methods is their reliance on force-matching alone, which can lead to energy predictions that differ by an arbitrary constant and are therefore unsuitable for absolute free energy comparisons. Here, we introduce a novel methodology with a graph neural network (GNN)-based implicit solvent model, dubbed $\lambda$-Solvation Neural Network (LSNN). In addition to force-matching, this network was trained to match the derivatives of alchemical variables, ensuring that solvation free energies can be meaningfully compared across chemical species. Trained on a dataset of approximately 300,000 small molecules, LSNN achieves free energy predictions with accuracy comparable to explicit-solvent alchemical simulations, while offering a computational speedup and establishing a foundational framework for future applications in drug discovery.
\end{abstract}

\section{Background}

Molecular dynamics (MD) simulations are extensively used to study the structure and function of biological systems and estimate the free energy of protein-ligand binding \cite{Hansson2002, Hollingsworth2018}. This latter application is particularly important in computer-aided drug discovery as accurate binding affinity estimations for potential drug candidates help reduce costs and accelerate early-stage drug development. However, achieving comprehensive conformational sampling requires sufficiently long simulations; the immense computational cost of free energy calculations limits the comprehensive analysis to only a few ligands despite the need to screen millions of drug candidates \cite{Hollingsworth2018, Mao2023}.

One key factor affecting conformational sampling in simulations is the solvation effect\cite{Anandakrishnan2015}. Traditional explicit solvent models simulate solvent molecules in a bulk solution surrounding the solute, offering highly accurate sampling but at a substantial computational cost. Implicit solvent models provide a faster alternative by replacing discrete molecular interactions, formed by each solvent molecule, with a mathematical approximation of the mean forces throughout the solution \cite{Kleinjung2014}. Thus, implicit solvation reduces simulation time by eliminating the need to simulate the conformations of a large number of solvent molecules. However, implicit solvent models often lack accuracy in describing processes where the conformation of the solvent molecules is important. Due to the importance of protein-solvent interactions, the accuracy of protein-related free energy simulations still strongly depends on the chosen model; multiple attempts have been made to improve the accuracy of calculations for problems like protein folding\cite{Kleinjung2014, Gruebele2002}.In traditional implicit solvent frameworks, the total solvation free energy can be expressed as the sum of the polar (Coulombic) and non-polar (Lennard–Jones) contributions:

\begin{equation}
\Delta G = \Delta G_{\text{Coulomb}} + \Delta G_{\text{LJ}}
\end{equation}

For many implicit solvent models, the solvation free energy is approximated as:

\begin{equation}
\Delta G \approx \Delta G_{\text{GB}} + \Delta G_{\text{SASA}}
\end{equation}
Here, $\Delta G_{GB}$ represents the polar contribution, while $\Delta G_{SASA}$ accounts for the non-polar component. The polar term can often be accurately estimated using methods such as generalized Born (GB) or Poisson-Boltzmann (PB). However, both GBSA (Generalized-Born Surface Area) and PBSA (Poisson–Boltzmann Surface Area) models use a simple solvent-accessible surface area (SASA) term to model the nonpolar contribution to free energy, a simplification that is prone to significant errors \cite{Huang2018}.

 More recently, new machine learning (ML) modeling techniques have been employed for implicit solvent descriptions, where neural network potentials allow for accurate calculations of solvation forces for any given conformation\cite{Cumberworth2015, Katzberger2023, Yao2023, Coste2023, Arsiccio2022}. Additionally, they offer transferable potentials across chemical space, enabling generalization to a wide range of molecular environments and providing accurate estimations of chemical properties and structural ensembles. 
 
 Such neural networks are typically trained using the force-matching approach, in which model parameters are optimized to minimize the discrepancy between forces predicted by the network and reference forces obtained from the differentiation of interatomic potentials with respect to spacial coordinates. This approach, however, determines potential energies only up to an arbitrary constant, limiting their utility for free energy simulations. The standard loss function for force-matching is defined as follows, where the model prediction is denoted by $f$:
 \begin{equation}
\mathcal{L} = 
\left(
\left\langle \frac{\partial U_{\text{solv}}}{\partial \mathbf{r}_i} \right\rangle - \frac{\partial f}{\partial \mathbf{r}_i} \right)^2
\end{equation}

While this methodology is appropriate for conformational calculations for the ligand, solvent-induced conformational variability complicates training. Proper training typically involves averaging forces over these solvent boxes to estimate Mean Applied Forces (MAFs) on solute atoms. The integration of the MAFs would have to account for solute conformation influenced by differing solvent landscapes, which is inherently the solvation free energy itself or the potential of mean forces (PMF). Hence, models trained on them perform well in capturing conformational landscapes but poorly in predicting absolute free energies. This complication restricts ML-based implicit solvent models to the calculation of forces or "force-matching" \cite{Chen2021}. 

In this paper, we advance the force-matching approach by extending the training procedure to incorporate additional energy derivatives, specifically electrostatic and steric coupling factors, and train a comprehensive neural network capable of simultaneously sampling conformational space and providing accurate potential of mean force (PMF) values. Additionally, our method computes solvation energies with significantly improved accuracy and simulation stability compared to implicit solvent models, while being substantially faster than explicit TIP3P calculations \cite{Jorgensen1983}.

\section{Methods}

Explicit solvent models are currently considered the gold standard for calculating solvation energies, but their major drawback is the high computational cost associated with long simulation times \cite{Zhang2017}.Analytical implicit solvent methods, such as the Poisson–Boltzmann or Generalized Born approaches, treat the solvent as a continuum medium and can accurately capture long-range electrostatic interactions, although they are less effective at describing local solvation effects. Recently, machine learning–based approaches have shown promise in surpassing implicit solvent models in accurately capturing these local effects. However, most current ML models are trained on data generated from explicit solvent simulations, as experimental solvation data are difficult to obtain. Consequently, their accuracy and generalizability remain constrained by the limitations of the underlying atomistic simulations used for training \cite{Chen2021}. Our LSNN approach, building on the foundational work of \citeauthor{Katzberger2023}, achieves near–explicit-solvent accuracy while maintaining computational efficiency comparable to implicit solvent models. The \citeauthor{Katzberger2023} model provides a strong basis for predicting conformational landscapes using a three-layer invariant Graph Neural Network (GNN) trained with standard GBSA parameters (GBSAGBn2Force and SASA) \cite{Katzberger2023}. In our implementation, the nonpolar solvation contribution predicted by the GNN is combined with the estimated polar component: 

\begin{equation}
\Delta G_{\text{non-polar}} = \sum_{i=1}^{N} \sigma \left( \phi(R, q, r_s, r, R_{\text{cutoff}}) \right) \gamma (r + r_w)^2
\end{equation}

where, $\sigma$ is a sigmoid function, $\phi$ is representing the GNN, and $R, q, r_s, r, R_{\text{cutoff}}$ are the GBn2 parameters, charges, and other atomistic representations. More information on the model can be found in the original paper \citeauthor{Katzberger2023}. 

While this methodology is appropriate for conformational calculations for the ligand, it continues to be restricted for energy calculations. To overcome this limitation, we extend the model by incorporating additional derivatives of the solvation energy, namely the electrostatic and steric coupling factors, $\lambda_{sterics}$ and $\lambda_{elec}$. These coupling factors scale the interaction energy calculated by the Coloumbic function and soft-core Lennard-Jones energy function \cite{Beutler1994, Kong1996LambdaDynamics}. As the primary variables in calculating $\Delta G_{solv}$ are the ligand conformations and interaction forces, these derivatives are scaled appropriately and used to construct a conservative vector field, allowing the calculated scalar potential $f$ to better approximate the true Potential of Mean Force (PMF).

Our modified MSE loss function is thus: 
\begin{equation}
\mathcal{L} = 
w_F\left(
\left\langle \frac{\partial U_{\text{solv}}}{\partial \mathbf{r}_i} \right\rangle - \frac{\partial f}{\partial \mathbf{r}_i} \right)^2
+
w_{_{\text{elec}}}\left(
\left\langle \frac{\partial U_{\text{solv}}}{ \partial \lambda_{\text{elec}}} \right\rangle - \frac{\partial f}{\partial \lambda_{\text{elec}}} \right)^2
+
w_{_{\text{steric}}}\left(
\left\langle \frac{\partial U_{\text{solv}}}{ \partial \lambda_{\text{steric}}} \right\rangle - \frac{\partial f}{\partial \lambda_{\text{steric}}} \right)^2
\end{equation}
where, $w_F$, $w_{elec}$, and $w_{steric}$ are empirically tuned weights for each term. This loss encourages the model to match the changes in the true solvation potential with respect to the interaction factors alongside the MAFs. Consequentially, nonpolar contributions for the model can be updated as follows:

\begin{equation}
\Delta G_{\text{non-polar}} = \sum_{i=1}^{N} 
\sigma \left( 
\phi(R, \lambda_{\text{elec}} q, \lambda_{\text{sterics}} r_a, r, R_{\text{cutoff}}) 
\right) 
\gamma_{\lambda_{\text{sterics}}} (r_i + r_w)^2
\end{equation}
We augment the GNN model by incorporating steric and electrostatic scaling factors ($\lambda$ values), which individually account for Coulombic interactions and Lennard-Jones potential.  Derivatives can be calculated with respect to $\lambda$ values to calculate the energy derivatives. Electrostatic calculations are approached similarly, without the additional SASA evaluation. 
\begin{figure}
    \centering
    \includegraphics[width=1\linewidth]{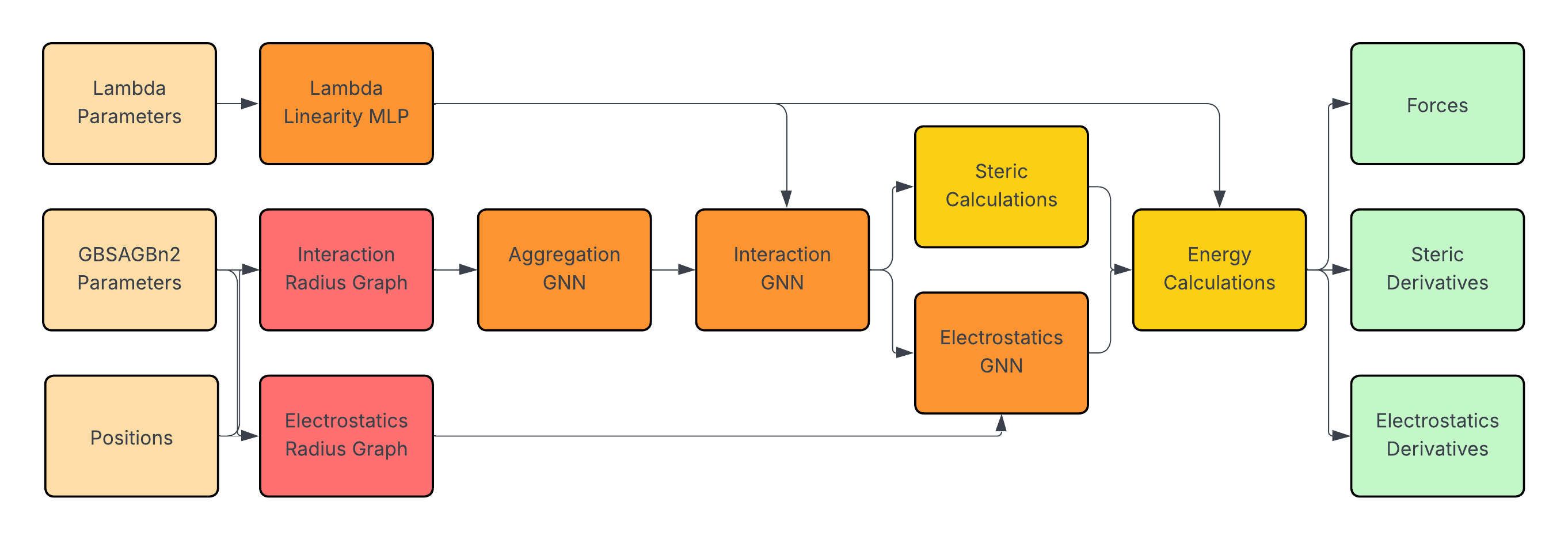}
    \caption{LSNN framework. This GNN model was modified from \citeauthor{Katzberger2023} to enable modelling of alchemically-modified ligands.}
    \label{fig:enter-label}
\end{figure}

The interaction GNN is integrated into the LSNN architecture as shown in Figure 1. Because the influence of $\lambda$ values, especially $\lambda_{\text{ster}}$, on energy derivatives is nonlinear, an MLP is used to transform them into values that are linearly related to the final energy function. While steric energies are harder to predict accurately, electrostatic components are even more difficult to generalize due to their higher interaction density. To address this, an additional GNN with a larger interaction radius is used. Each energy term is then multiplied by its corresponding $\lambda$ value to ensure the total energy is zero in fully decoupled states, preserving proper scaling for accurate PMF calculations. Model derivatives are computed via Pytorch Autograd with weights $w_F$, $w_{sterics}$, and $w_{elec}$ applied in a 1:1:1.2 ratio, respectively \cite{Paszke2019}.

To ensure broad chemical diversity for generalizing estimations, we selected a subset of approximately 280K small neutral molecules from the BigBind dataset \cite{Brocidiacono2023}. The data was split into 80:10:10 train, validation, and test sets, ensuring that any molecule within 0.3 Tanimoto similarity (with 2048-bit radius 3 Morgan fingerprints) to any molecule in Freesolv was in the test set \cite{Mobley2014, Morgan1965}. MAFs and interaction derivatives were computed on frozen conformations using OpenMM with the GAFF-generated force field for 0.5 ns at a 2 fs step size\cite{Eastman2023, Wang2004}. To generalize the PMF function, all energy derivatives were computed by scaling $\lambda_{sterics}$ and $\lambda_{elec}$ uniformly across allowed values to ensure steric coupling was never minimized independently of electrostatic annihilation. Alchemically modified MD simulations calculations were conducted using the AbsSolv framework, and implicit solvent energy calculations were carried out using OpenMMTools \cite{openmmtools, BoothroydAbsolv}. For derivative calculations, finite difference approximations were used throughout, with an empirically optimized $\epsilon$ value of $10^{-7}$.

The total simulation time was 17 days on 20 L40 GPUs. The additional collation of the dataset took $\sim$1 day on 1 L40 GPU. Model was trained for 75 Epochs with a batch size of 100, a learning rate (LR) of 1e-3, and gradient clipping 5.0. Training time was 7 minutes per epoch and $\sim$2 hours in totality on 1 L40 GPU.

\section{Results}

\subsection{Benchmarking Conditions}

We selected OBC2 and GBn2 as representative implicit solvent models as they are currently the most popular methods alongside TIP3P for explicit solvent\cite{Hong2013, Totrov2004, Jorgensen1983}. To properly test the efficacy of LSNN against traditional and computational models, we first needed to find the optimal amount of lambda windows and step size for all models for solvation energy calculations using Multistate Bennett acceptance (MBAR)\cite{Klimovich2015}. MBAR allows us to examine a wide range of lambda windows and estimate the free energy changes using their phase space overlap. We performed all tests on 647 neutral small molecules with experimental hydration free energies from the FreeSolv dataset \cite{Mobley2014}. Although previous studies have documented the differences between existing implicit and explicit solvation models, our main goal was to implement a method that can realistically represent the solvent faster and more accurately than existing models. To fully envision this, benchmarking was conducted to calculate varied simulation parameters to achieve peak computational performance for each solvent model. For TIP3P, we applied the same parameters as in the FreeSolv benchmark: 20 $\lambda$ windows with 0.5 ns simulations per window \cite{Mobley2014}. 

Although implicit solvent models may require more computation time per simulation frame, the overall efficiency comes from replacing costly solvent-solute dynamics with analytical estimates, ulitimately reducing variability. This efficiency also means that adequate sampling can often be achieved with fewer intermediate $\lambda$ states than in explicit solvent simulations \cite{Klimovich2015}. Once the phase-space overlap among adjacent $\lambda$ windows is sufficient, adding more $\lambda$ states yields negligible accuracy gains. To identify the minimum required states, we compared results from 20 $\lambda$ reference states obtained across all test molecules using averaged MBAR overlap matrices for all 3 implicit solvents. In GBn2 and LSNN, the phase space overlap remains stable across the fully coupled, fully decoupled, and intermediate $\lambda$ states, indicating consistent sampling.In contrast, OBC2 exhibited a small deviation in electrostatic overlap, accounted for by an additional electrostatic window at $ \lambda_{elec} = 0.5$.

\captionsetup[subfigure]{skip=2pt}

\begin{figure}[h!]
    \centering
    \begin{subfigure}[b]{0.5\linewidth}
        \centering
        \includegraphics[width=\linewidth]{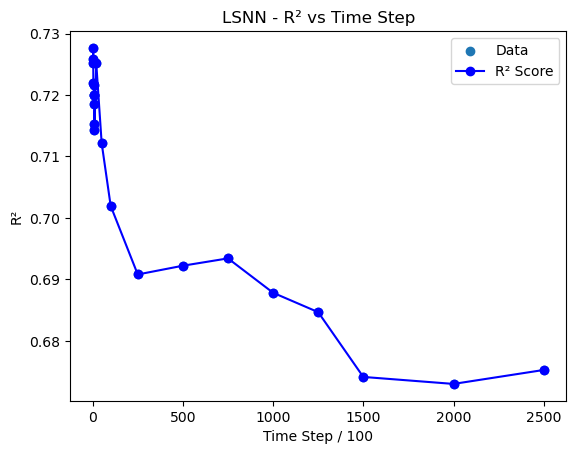}
        \caption{LSNN}
        \label{fig:top}
    \end{subfigure}

    \vspace{0.5em}

    \begin{subfigure}[b]{0.48\linewidth}
        \centering
        \includegraphics[width=\linewidth]{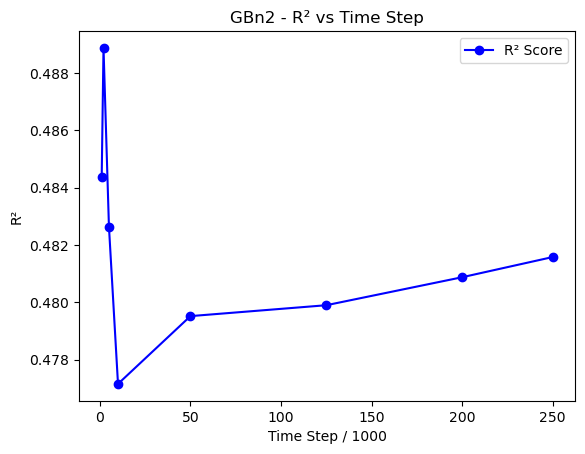}
        \caption{GBn2}
        \label{fig:bottom-left}
    \end{subfigure}
    \hspace{0.02\linewidth}
    \begin{subfigure}[b]{0.48\linewidth}
        \centering
        \includegraphics[width=\linewidth]{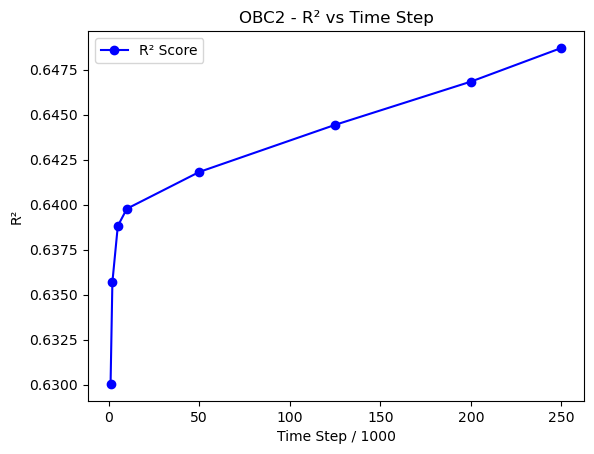}
        \caption{OBC2}
        \label{fig:bottom-right}
    \end{subfigure}

    \caption{Solvation Energy Accuracy Benchmark against Simulation Times for Implicit Solvent Models. GBN2 and OBC2 display similar trends while LSNN fluctuates. Result calculations were conducted with optimal speed and accuracy. }
    \label{fig:composite}
\end{figure}

While the impact of increasing intermediate states is minimal once proper MBAR overlap is established, successive conformational sampling of molecules can still cause fluctuations in the R² value. Initial test simulations were conducted with 250,000 frames using 2 fs time steps, resulting in a total simulation time of 0.5 ns. All other variables remained consistent with dataset generation. Figure 2 illustrates the changes in R² for each solvent model as the sampling window is reduced. 

For LSNN, the R² curve peaks at around 0.6 ps and then steadily decreases. This decline may result from numerical inaccuracies, such as divergence during electrostatic derivative calculations or general model instability. While such effects can impact conformational calculations, they are \textit{not currently} a major concern for the current LSNN framework. As solvation continuums have limited variability, shorter simulations are typically utilized regardless. Since LSNN is currently only trained for small molecules, the shorter simulation bottleneck is not significant. However, as the framework is expanded to larger molecules, additional MAFs from equilibrated and non-equilibrated conformers must be trained. 

A similar trend is observed by GBn2 which maximizes at 4ps. However, the GBn2  seems to increase in accuracy beyond 200,000 timesteps or 0.4 ns. The practicality past this point may be questioned as GBn2 as computational efficiency is drastically outweighed.  While the trends displayed for GBn2 are unexplained, the similar trend in LSNN can possibly also be attributed to utilizing GBn2 parameters within the model. Future iterations of LSNN will utilize OBC2 parameters instead. In contrast, OBC2 resembles a graph where additional simulation time always leads to higher accuracy, as expected. Based on these results, we selected a simulation time of 10 ps as the optimal balance between speed and accuracy for OBC2.

For benchmarking, Alchemlyb was used to decorrelate and remove uncorrelated simulation frames, compute binding affinities with MBAR, and generate MBAR overlap matrices with the appropriate $u_nk$ values \cite{Wu2024}. All simulation calculations were performed using similar scripts (available on GitHub) with the optimal hyperparameters outlined above. Any samples yielding NaN values during simulations or exceeding large $\Delta$G values ( >|100| kcals) were excluded from results and automatically considered as failures.

\subsection{Benchmarking Results}

\begin{figure}[H]
    \centering
    \includegraphics[width=0.5\linewidth]{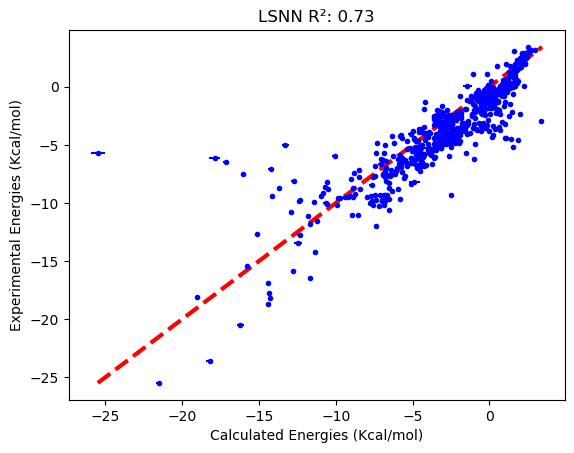}
    \caption{Accuracy of LSNN compared against experimentally calculated values}
    \label{fig:lsnn_accuracy}
\end{figure}

According to Figure~\ref{fig:lsnn_accuracy}, calculations for LSNN were successful for 638 out of 647 compounds tested during experimentation. Overall, LSNN had a high correlation with experimentally tested compounds, achieving an R\textsuperscript{2} of 0.73. Inaccuracies can also be formed due to insufficient phase space overlap. 

\FloatBarrier

In Figure 4, TIP3P calculations were, as expected, more accurate and were more successful for 646 out of 647 compounds during experimentation— one compound led to nonconverging energy minimization. The R\textsuperscript{2} reflected for $\Delta$G is 0.86.

%

\begin{figure}[H]
    \centering
    \includegraphics[width=0.5\linewidth]{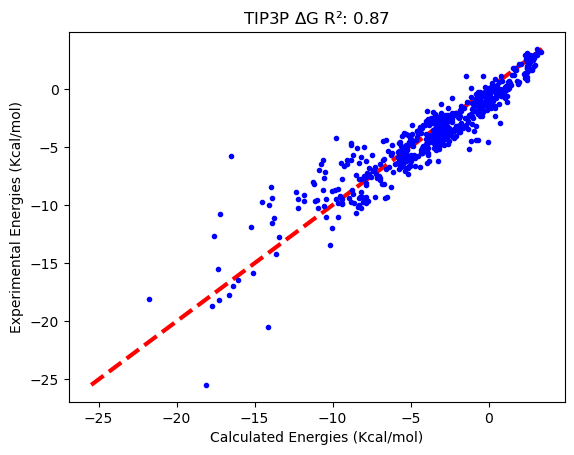}
    \caption{Accuracy of TIP3P compared against experimentally calculated values}
    \label{fig:image8}
\end{figure}

\FloatBarrier

Figure 5 compares GBn2 and OBC2, both which shows a lower accuracy, with R\textsuperscript{2} of 0.48 and 0.63 respectively, and lower success rates—610 and 611 out of 647 compounds. Similar to LSNN, calculation accuracy declined sharply as the binding affinity deviated from zero. Some inaccuracies were consistent between the LSNN model and the implicit solvent model, particularly among the outliers. These errors likely stem from inaccuracies in the implicit solvent parameters estimations themselves.

\begin{figure}[htbp]
    \centering
    \begin{subfigure}[b]{0.48\linewidth}
        \centering
        \includegraphics[width=\linewidth]{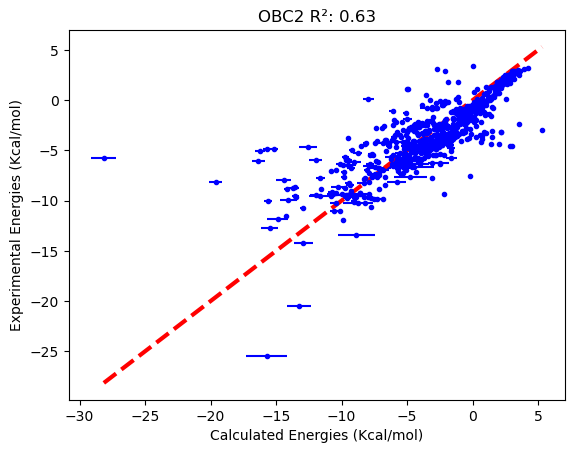}
        \caption{OBC2}
        \label{fig:image8}
    \end{subfigure}
    \hfill
    \begin{subfigure}[b]{0.48\linewidth}
        \centering
        \includegraphics[width=\linewidth]{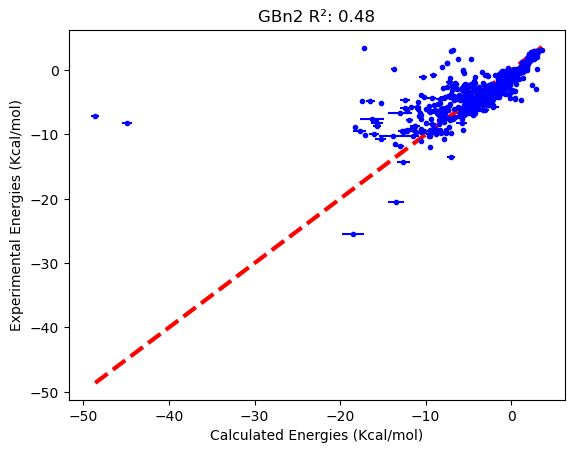}
        \caption{GBn2}
        \label{fig:image9}
    \end{subfigure}
    \caption{Calculated solvation energies for conventional implicit solvent models.}
    \label{fig:side-by-side}
\end{figure}

In Table 2, the cumulative results are displayed. Machine learning algorithms tend to be slower than other mathematical approaches, but LSNN tends to compute solvation energies faster (20.47 seconds) due to a significantly shorter simulation time than OBC2 (21.81 seconds) and remains comparable to GBn2 (15.82 seconds). This statistic aligns with computations as LSNN relies on GBn2 parameters. Moreover, LSNN remains \textit{signficantly} faster than TIP3P (1658.54 seconds or 27.64 minutes), where the cost of accuracy is proportionately less. While the practicality of OBC2 and GBN2 can be questioned, LSNN poses an effective balance in accuracy while maintaining speeds comparable to other mathematical approximations.  

\begin{table}[h!]
\centering
\begin{tabular}{@{}lSS@{}}
\toprule
\textbf{Method} & \textbf{Speed (seconds)} & \textbf{Accuracy (R\textsuperscript{2})} \\
\midrule
TIP3P - $\Delta$G  & 1658.54 & \textbf{0.87} \\
\textbf{LSNN}      & 20.47 & 0.73 \\
OBC2              & 21.81          & 0.63 \\
GBn2              & \textbf{15.82}          & 0.48 \\
\bottomrule
\end{tabular}
\caption{Comparison of methods based on speed and accuracy.}
\end{table}

\subsection{Preliminary Binding Affinity Results}

While the current LSNN model is trained exclusively on small, neutral molecules, our future direction involves generalizing to include larger biomolecular systems, such as peptides, proteins, and protein-ligand complexes. Notably, Katzberger's original model showcased scalability to small peptides. To explore this application, we performed preliminary binding free energy calculations on a protein-ligand complex utilizing Molecular Mechanics - Generalized Born Surface Area (MM-GBSA) \cite{Genheden2015}. 

\begin{equation}
\Delta G_{\mathrm{bind}} = G_{\mathrm{complex}} - \left( G_{\mathrm{protein}} + G_{\mathrm{ligand}} \right)
\end{equation}

\begin{equation}
\Delta G_{\mathrm{bind}} = \Delta E_{\mathrm{MM}} + \Delta G_{\mathrm{solv}} \quad \text{(entropy term } T\Delta S \text{ omitted)}
\end{equation}

\begin{figure}[h!]
    \centering

    \begin{subfigure}[b]{0.48\linewidth}
        \centering
        \includegraphics[width=\linewidth]{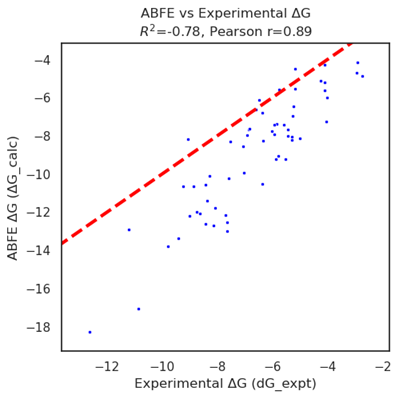}
        \caption{ABFE}
        \label{fig:top-left}
    \end{subfigure}
    \hspace{0.02\linewidth}
    \begin{subfigure}[b]{0.48\linewidth}
        \centering
        \includegraphics[width=\linewidth]{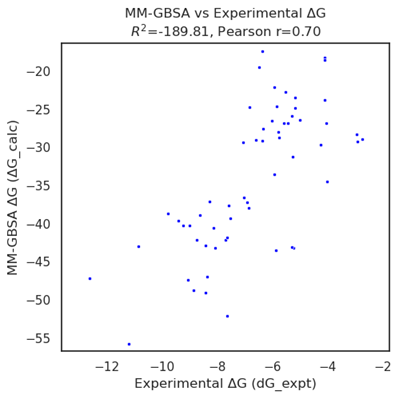}
        \caption{MM-GBSA}
        \label{fig:top-right}
    \end{subfigure}

    \vspace{0.5em}

    \begin{subfigure}[b]{0.48\linewidth}
        \centering
        \includegraphics[width=\linewidth]{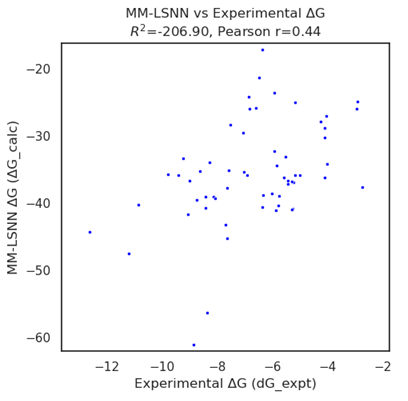}
        \caption{MM-LSNN}
        \label{fig:bottom-left}
    \end{subfigure}
    \hspace{0.02\linewidth}
    \begin{subfigure}[b]{0.48\linewidth}
        \centering
        \includegraphics[width=\linewidth]{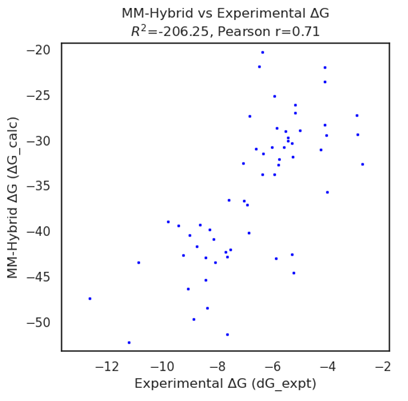}
        \caption{MM-Hybrid}
        \label{fig:bottom-right}
    \end{subfigure}

    \caption{Accuracy Benchmark of Common Binding Affinity Calculations against LSNN}
    \label{fig:composite}
\end{figure}

By substituting free energy estimations formed by GBSA with the PMFs calculated by LSNN, we can yield a modified methodology: MM-LSNN. Alongside these calculations, a hybrid MM-GBSA/LSNN approach where $\Delta G_{\mathrm{lig}}$ was calculated using LSNN is also tested. 

Alongside the implicit solvent calculations that were performed on the same trajectory, Absolute Binding Free Energy (ABFE) calculations on the same molecules are also performed. The experimental and ABFE results were precomputed by \cite{Alibay2022}. Note that all simulations for the implicit solvent model were completed utilizing OBC2, and only energy calculations were done with the respective models. 

In Figure 6, as expected, LSNN's standalone performance is limited to full protein systems due to its training domain with a lower R of 0.44, and the hybrid model shows only marginal improvements over MM-GBSA (R of 0.71 vs 0.70) due to the ligand's desolvation contributing minimally to the total binding energy. The calculated binding free energy of MM-LSNN generally is more negative than MM-GBSA as well, leading to a lower $R^2$. We estimate that the overly dense graph used to train LSNN led to an overestimation of long-range interactions in the calculation of $\Delta G_{\mathrm{complex}}$ and $\Delta G_{\mathrm{protein}}$. This hypothesis is further evident by the narrower range found in MM-Hybrid compared to MM-GBSA and MM-LSNN. Importantly, despite not being explicitly optimized for binding energy calculations, both LSNN-based approaches exhibited a clear and consistent linear correlation with experimental values, highlighting their strong potential for further optimization and generalization to larger biomolecular systems.

\section{Discussion}

LSNN is significantly faster than traditional explicit solvent while maintaining competitive accuracy. While LSNN's foundational work validate the core algorithmic foundation, it also demonstrates several extensions for future research. LSNN provides a faster and more stable model for computing the solvation continuum, making it highly suitable for accurate solvation calculations and substantially accelerating drug discovery. 

The current limitations of LSNN for smaller molecules are the lack of proper generalization and deviations for conformational ensembles beyond the minimized state. Although LSNN is optimized for thermodynamically consistent free energy calculations rather than comprehensive conformational sampling, recent work has shown that GNN-based implicit solvent models can capture proper sampling within solvents \cite{Katzberger2025}. As these errors led to a certain level of inaccuracy in free energy calculations, future iterations of LSNN will address these limitations by incorporating volumetric terms to enable more comprehensive enthalpic estimations and extending training to non-equilibrated and diverse sampled conformational states. Moreover, the framework provided can be extended beyond solvation energy calculations and applied to binding free energy calculations. Future efforts will expand the training dataset to include charged ligands and optimize the model architecture to reduce computational complexity for numerous residues by investigating sparse graph representations to handle the increased complexity of protein systems and subgraph representations of residues to handle localized interactions. More importantly, LSNN will attempt to form a more holistic hydration solvent.

\section{Conclusion}
LSNN successfully changes force-matching, changing the convention of ML-based transferable potentials. We capture ligand desolvation trends and maintain consistent ordering across diverse ligands. Validation of the framework established in this paper enables calculation of energy functions and can be extended to applications such as estimation of the energy of protein-ligand interaction. Improvements to LSNN will continue to expand its current and future usages.

%

\bibliography{achemso-demo}

\end{document}